\documentclass{article}
\usepackage{amssymb}
\usepackage{epsfig}
\usepackage{graphics}
\DeclareSymbolFont{AMSb}{U}{msb}{m}{n}
\DeclareMathSymbol{\N}{\mathbin}{AMSb}{"4E}
\DeclareMathSymbol{\Z}{\mathbin}{AMSb}{"5A}
\DeclareMathSymbol{\R}{\mathbin}{AMSb}{"52}
\DeclareMathSymbol{\Q}{\mathbin}{AMSb}{"51}
\DeclareMathSymbol{\C}{\mathbin}{AMSb}{"43}
\usepackage{epic}
\usepackage{eepic}
\usepackage{times}
\usepackage{color}

\newcommand{\dahntab}[1]{
  \newbox\mybok%
  \setbox\mybok=\hbox{\vbox{
      \begin{tabbing}
        #1
      \end{tabbing}%
    }}

  \newdimen\bokwidth%
  \bokwidth=\wd\mybok%
  \newdimen\myl%
  \myl=\textwidth%
  \divide\myl by 2%
  \divide\bokwidth by -2%
  \advance\myl by\bokwidth%
  \vrule width\myl height 0pt depth 0pt%
  \usebox\mybok%
}

\newenvironment{proof}{\begin{trivlist}\item[]{\emph
{Proof.}}}{\hfill \end{trivlist}}

\newcommand{\keyw}[1]{{\bf #1}}

\def\goesto{\rightarrow}

\newtheorem{lemma}{Lemma}[section]
\newtheorem{theorem}{Theorem}[section]
\newtheorem{definition}{Definition}
\newtheorem{proposition}{Proposition}

\newcommand{\implies}{\Rightarrow}
\newcommand{\cminus}{\dot{-}}
\newcommand{\restored}{{\sc Restored}}

\def\BibTeX{{\rm B\kern-.05em{\sc i\kern-.025em b}\kern-.08em
    T\kern-.1667em\lower.7ex\hbox{E}\kern-.125emX}}
\def\AND{\wedge}
\def\OR{\vee}
\newcommand{\qed}{\hfill \ensuremath{\Box}}

\begin{document}

\title{Counting Preimages of TCP Reordering Patterns}

\author{Anders Hansson, \\ Discrete Simulation Sciences (CCS-5),
\\Los Alamos National Laboratory\\P.O.\ Box 1663, Mail Stop M997,\\
Los Alamos, NM 87545, USA. \\ hansson@lanl.gov \and
Gabriel Istrate\footnote{Corresponding author}, \\
eAustria Research Institute, \\ Bd. Corneliu Coposu no. 4, cam. 045B\\
Timi\c{s}oara, RO-300223, Romania. \\ gabrielistrate@acm.org}

\maketitle

\begin{abstract}
Packet reordering is an important property of network traffic that
should be captured by analytical models of the Transmission Control
Protocol (TCP). We study a combinatorial problem motivated by
\restored ~\cite{restored}, a TCP modeling methodology that
incorporates information about packet dynamics. A significant
component of this model is a many-to-one mapping $B$ that transforms
sequences of packet IDs into {\em buffer sequences} in a manner that
is compatible with TCP semantics. We show that the following hold:

\begin{itemize}

\item There exists a linear time algorithm that, given a buffer
sequence $W$ of length $n$, decides whether there exists a permutation
$A$ of $\{1,2,\ldots, n\}$ such that $A\in B^{-1}(W)$ (and constructs
such a permutation, when it exists).

\item The problem of counting the number of permutations in
$B^{-1}(W)$ has a polynomial time algorithm.

\item We also show how to extend these results to sequences of IDs
that contain repeated packets.

\end{itemize}

\end{abstract}


\section{Introduction}

Consider a sequence of {\em TCP packets}, identified by their integer
IDs, as handled by their {\em receiver}. The receiver must forward the
packet sequence to an {\em application}, subject to respecting {\em
packet sequence integrity}.  That is, at every moment the IDs of
packets forwarded to the application must form a contiguous sequence
$1,2,\ldots, m$, for some $m\geq 1$. Packets can arrive out-of-order
and thus need to be buffered. Several copies of a packet can arrive,
but only one copy of a given packet is useful (and will be stored, if
needed). We assume that the receiver evicts a given packet from the
buffer and passes it to the application as soon as possible, i.e., as
soon as the packet sequence integrity constraint is satisfied.

A given sequence $A=(A_{1}, \ldots, A_{n})$ of packet IDs yields a
corresponding sequence $B(A)=(B_{A,1},\ldots , B_{A,n})$ representing
the evolution of the buffer size. In this paper we are interested in
the following problem: given a sequence of positive integers $W$, what
is the complexity of

\begin{enumerate}

\item Deciding whether there exists a permutation $A$ with $W=B(A)$?

\item Counting the number of permutations in the set $B^{-1}(W)$?

\end{enumerate}

\section{Motivation}

The problem we described in the introduction arises in the context of
analytical modeling of TCP dynamics. Therefore, the reader only
interested in the combinatorial aspects of the problem can focus on
the remaining sections. This section explains in detail the motivation
for the problem.

While a lot of attention has been given to modeling the temporal
aspects of TCP traffic (see e.g.\ Jaiswal \emph{et al}.\
\cite{jaiswal-inferring-infocom}), the dynamics of packet IDs has not
received the same attention. As Bennett \emph{et al}.\
\cite{pathological} have shown, packet reordering is more widespread
than originally believed, and is increasingly becoming so, due to
technological advances such as link striping and mobile
communications. Packet reordering has many severe effects on overall
traffic characteristics, hence it is an important component of TCP
dynamics (we refer the reader to \cite{pathological} for further
discussion).

Paper~\cite{restored} introduced \restored, a methodology for semantic
compression and regeneration of large TCP traces. \restored\ is based
on the following observation: TCP guarantees to deliver an ordered
packet stream to the application layer and needs to buffer packets
that arrive out-of-order. Consequently, the received packets can be
classified into two types: those that could be immediately passed to
the application layer, and those that have to be temporarily
buffered. A received packet that allows the buffer to flush is called
a {\em pivot packet}. All packets appearing in order are trivially
pivots. \restored\ divides the received sequence into segments,
bounded by pivot packets. Segments correspond to one of two {\em
phases}:

\begin{itemize}

\item An {\em ordered phase}, in which no reordering is present, thus
there is no need for buffering.

\item An {\em unordered phase}, in which there is reordering and
buffering.\footnote{A technical assumption we will employ is that
duplicates of packets that have already been uploaded to the
application layer are discarded. This is a sensible assumption, given
TCP behavior.} Each occurrence of this phase ends when a pivot packet
is received.

\end{itemize}

\restored\ preserves packet reordering properties of TCP traffic, up
to a notion of semantic equivalence of packet traces. This notion is
called {\em behavioral equivalence} and can be motivated as follows:

\begin{definition}
Let $ACK_{i}$ be defined as the smallest integer that does not appear
among the first $i$ packet IDs (also, define $ACK_{0}=1$). Parameter
$ACK_{i}$ is called {\em the acknowledgement (ACK) at stage $i$}.
\end{definition}

The previous definition relies on the simplifying assumption that in
the implementation of TCP each received packet is ACKed, and that
value $ACK_{i}$ is the only information carried by the ACK packet. Of
course, real-life acknowledgment policies of TCP can be more
complicated \cite{stevens-tcp-1}.

Consider now the following two packet ID sequences: 4 2 3 1 and 4 3 2
1.

Both these sequences trigger identical ACK responses, namely 1 1 1 5,
i.e., we arrive at the following two mappings:
\begin{center}
\begin{equation}\label{eq}
\begin{array}{rrrrcrrrl}
4 & 2 & 3 & 1 & \goesto & 1 & 1 & 1 & 5, \\
4 & 3 & 2 & 1 & \goesto & 1 & 1 & 1 & 5.
\end{array}
\end{equation}
\end{center}

Since TCP is a {\em receiver-driven} protocol, assuming identical
network conditions, and discounting possible differences in the value
of the congestion window at the beginning of the sequences, {\em the
two ID sequences trigger identical responses from the receiver, and
should thus be regarded as indistinguishable from the standpoint of
TCP dynamics}.

\begin{definition}
Two sequences of packets $P$ and $Q$ are {\em behaviorally equivalent}
(written $P\equiv_{beh} Q$) if they lead to the same sequences of
ACKs.
\end{definition}

In practice one might want a notion of equivalence that is even more
restrictive than behavioral equivalence. This was, for instance, the
case of \restored. Its original motivation was to provide a way to
compress TCP traces and estimate various measures of quality of
service of the original traces by reconstructing ``compatible''
sequences. Many measures of packet reordering have been proposed in
the networking literature
\cite{reordering-lcn02,reordering-ifip05,ietf-draft-reordering}. Given
such a measure $M$, one way to guarantee that sequences produced by
\restored \ resemble the original sequence with respect to measure $M$
is:
\begin{enumerate}

\item Identify an equivalence notion of ID sequences $\equiv$ such
that {\em $M$ is consistent with respect to $\equiv$}, that is
\begin{equation} \label{equiv-eq}
(\forall A,B)\mbox{: }(A\equiv B)\implies (M(A)=M(B)).
\end{equation}

\item Make sure that for any sequence $A$, the sequence $R(A)$
regenerated by \restored \ satisfies $R(A)\equiv A$.

\end{enumerate}
(See also \cite{restored-metrics} for more discussion and
clarification). Behavioral equivalence might be too coarse (as an
equivalence relation) to guarantee consistency of many reordering
metrics and, thus, needs to be refined. In a companion paper
\cite{restored-shiva} we have considered such an equivalence notion,
based on the following notion of buffer size:

\begin{definition}
Let $A=\{A_1,A_2,\ldots,A_n\}$ be a sequence of packet IDs. We define
the $\textup{FB}$ as an operator that after receiving a packet $A_i$
at time index $i$, outputs the difference between the highest ID
($H_i$) seen so far and the highest ID ($L_i$) that could be uploaded.
\begin{equation}
\textup{FB}(A_i) = H_i - L_i. \label{eq1}
\end{equation}
In other words, $\textup{FB}$ is the size of the smallest buffer
large enough to store all packets that arrive out-of-order, where the
definition of size accounts for reserving space for unreceived packets
with intermediate IDs as well. The \emph{buffer sequence}
$\textup{FB}(P)$ associated with a sequence $P$ of packet IDs is
simply a time-series of $\textup{FB}$ values computed after each
packet has been received.

Two sequences of packet IDs $P$ and $Q$ are {\em FB equivalent}
(written $P\equiv_{\footnotesize\textup{FB}} Q$) if
$\textup{FB}(P)=\textup{FB}(Q)$.
\end{definition}

This definition is directly related to the semantics of TCP, since it
preserves quantities such as the size of the AdvertisedWindow (see
\cite{peterson-davie}). Inverting the mapping $\textup{FB}$ can be
done in polynomial time \cite{restored-shiva}. However, the complexity
of computing the cardinality of the preimage $\textup{FB}^{-1}(W)$ was
left open, and was only solved in two special cases.

In this paper, we use a different notion, introduced below, for which
more precise results can be obtained.

\begin{definition}
Buffer size is the smallest size of a buffer that can store all
out-of-order packets. Two sequences of packets $P$ and $Q$ are {\em
buffer equivalent} (written $P\equiv_{buf} Q$) if $B(P)=B(Q)$, that is
the sequences of buffer sizes associated with receiving $P$ and $Q$
are identical.
\end{definition}

From a combinatorial perspective, buffer equivalence is more natural
than FB equivalence. Its relation with behavioral equivalence is,
however, slightly more complicated:

\begin{enumerate}
\item Buffer equivalence is {\em not} a refinement of behavioral
equivalence in general. Indeed, sequences of packet IDs 2 3 3 1 and 3
4 1 2 are buffer equivalent (they both map to sequence 1 2 2 0) but
{\em not} behaviorally equivalent (the ACKs are 1 1 1 4 and 1 1 2 5,
respectively). This stands in contrast to FB equivalence which is
indeed \cite{restored-shiva} a refinement of behavioral equivalence.
\item Buffer equivalence refines behavioral equivalence when
restricted to {\em permutations} (sequences with no repeats or lost
packets). For a formal statement and proof of this claim see
Proposition~\ref{refine} below.
\item Finally, buffer equivalence is incomparable (as an equivalence
notion) with FB equivalence \cite{restored-metrics}.
\end{enumerate}

On the other hand there exist reordering metrics $M$ defined in the
networking literature (e.g. {\em reorder buffer density}
\cite{ietf-reorder-density}) with the following properties:
\begin{enumerate}
\item $M$ only depends on packets received for the first time, and not on
repeat packets.
\item $M$ is inconsistent with respect to FB equivalence but consistent with
respect to buffer equivalence (metrics with opposite consistency properties
exist as well; see \cite{restored-metrics} for further details).
\end{enumerate}

The recovery of such metrics via the argument described in equation
~(\ref{equiv-eq}) motivates the problem we study in this note:
inverting the many-to-one map $B$ and counting the size of its
preimage. Results for map $B$ are slightly stronger than those proven
in \cite{restored-shiva} for map FB. Namely, computing the cardinality
of the preimage of map $B$, as well as returning one element from the
preimage can be done in polynomial time (even linear time for the
latter problem).

\section{Preliminaries}

We will use notation $x\cminus y=max\{x-y,0\}$.

We employ standard graph theoretic notions throughout. In this paper,
graphs are always bipartite and undirected. Denote by $d(v)$ the
degree of vertex $v$ and by $N(v)$ the set of neighbors of $v$.

\begin{definition}
A bipartite graph $G=(V_{1},V_{2},E)$ is {\em doubly convex} if there
exist permutations $\pi_{1}, \pi_{2}$ of vertex sets $V_{1},V_{2}$,
respectively, such that for every $i\in \{1,2\}$ and every vertex
$v\in V_{i}$ the set of vertices $w$ that are adjacent to $v$ forms an
interval (i.e. a set of consecutive nodes) of
$\pi_{3-i}(V_{3-i})$.
\end{definition}

\begin{definition}
A sequence of IDs $W$ is a {\em valid buffer pattern} if there
exists a permutation $A$ of $\{1,2, \ldots, |W|\}$ such that
$B(A)=W$.
\end{definition}

Note that any valid buffer pattern $W$ necessarily ends in a zero,
since for $A\in B^{-1}(W)$ all packets in $A$ can be passed to the
application layer when the last packet in $A$ is received. Also,
without loss of generality, one can assume that {\em the only}
position in a valid buffer pattern that is equal to zero is the last
one, since one can decompose a given pattern $W$ into disjoint
segments, bounded by those positions equal to zero (where the buffer,
therefore, gets flushed). To each such segment one can associate a
permutation of a contiguous set of IDs.

\section{Inverting Buffer Sequences}

Our main result is

\begin{theorem}\label{main}

The following are true:

\begin{enumerate}

\item There is an algorithm that, given an encoding $W =
W_1\#W_2\#\ldots\#W_n\#\#$ of a sequence of positive integers as
input (the $W_{i}$'s are integers in binary notation and $\#$ is a new
symbol) decides in time $O(|W|)$ whether $W$ is a valid buffer
pattern, and if this is the case constructs a permutation $A$ such
that $A=B^{-1}(W)$.

\item Counting the cardinality of the set of permutations in the
preimage $B^{-1}(W)$ can be done in polynomial time.

\end{enumerate}
\end{theorem}

\begin{proof}

We will provide, in essence, a reduction of the problem above to the
problem of finding a maximum matching in a special class of doubly
convex bipartite graphs \cite{convex-bipartite-acta}. The complexity
of this problem is linear in the number of vertices of the graph
\cite{convex-bipartite-acta}. Since the size of the bipartite graph
that is created by reduction is linear, the overall complexity of the
problem is linear.

A valid buffer sequence consists of positive integers, with the
exception of the last entry, which is zero. Any two consecutive values
of the buffer sequence $W_{i}$ and $W_{i+1}$ can only be in one of the
following situations:
\begin{enumerate}

\item $W_{i}=W_{i-1}+1$. This situation corresponds to one new
out-of-order packet being received at stage $i$. This holds for $i=1$
as well, if we let $W_{0}=0$.

\item $W_{i}<W_{i-1}$. This situation corresponds to the newly
received packet causing a non-empty portion of the buffer to be
flushed. In particular the ID of the received packet can be inferred
at this stage, and is equal to the smallest index of a packet not
received so far.

\item $W_{i}=W_{i-1}$. This situation corresponds to the packet
received at this stage being the first packet not previously received.
Receiving this packet does not cause any other packet to be sent to
the application layer.

\end{enumerate}

If the input sequence fails to satisfy these conditions (for
instance if there exists an index $i$ with $W_{i}-W_{i-1}>1$) then
the set of permutations in $B^{-1}(W)$ is empty. Otherwise, let
$S_{1}, S_{2}, S_{3}$ be the set of indices corresponding to the
three cases listed above.

During the course of the algorithm we will keep track of the value
$ACK_{i}$, computed assuming that $W$ is a valid buffer pattern.
 Initially $ACK_{0}=1$. We have the following recurrence relations
(mirroring the three cases described above):

\begin{enumerate}

\item The newly received packet is out-of-order. Thus, it does not
change the value of parameter $ACK$. Therefore
\begin{equation}\label{eq4}
ACK_{i}=ACK_{i-1}.
\end{equation}

\item The newly received packet has ID $ACK_{i-1}$. In addition, it
makes the buffer shrink in size from $W_{i-1}$ to $W_{i}$, which means
that
\begin{equation}
ACK_{i}=ACK_{i-1}+W_{i-1}-W_{i}+1.
\end{equation}

\item The newly received packet has index $ACK_{i-1}$ and does
not cause the buffer to shrink any more. Therefore
\begin{equation}\label{eq6}
ACK_{i}=ACK_{i-1}+1.
\end{equation}

\end{enumerate}

For all indices $i\in S_{2}\cup S_{3}$, the index of the received
packet is uniquely determined, and equal to $ACK_{i-1}$.

We will now create a bipartite graph $G=(V_{1}, V_{2},E)$. Nodes in
$V_{1}$ correspond to stage indices $i\in \{1,\ldots n\}$. Nodes in
$V_{2}$ will correspond to packet IDs.  First, let $V_{1}= S_{1}$,
and let $V_{2}=\{1,\ldots, n\}\setminus \{ACK_{i-1}|\mbox{ }i\in
S_{2}\cup S_{3}\}$. Clearly $|V_{2}|= n -|S_{2}\cup S_{3}|= |S_{1}|=
|V_{1}|$. Second, given node $i\in V_{1}$, add edges to all vertices
$j\in V_{2}$ such that $j>ACK_{i}$.

With this definition we have:

\begin{lemma}\label{biject}

Permutations from the set $B^{-1}(W)$ are in bijective correspondence with
elements of $MATCH(G)$,  the set of all perfect matchings in $G$. In
particular $B^{-1}(W)\neq \emptyset$ if and only if $G$ has a
perfect matching.
\end{lemma}
\begin{proof}

Each permutation can be seen as a set of pairs $(i,j)$. By the
previous discussion, the set of acknowledgements $\{ACK_{i}\}_{i\geq
0}$ is the same for any permutation in $B^{-1}(W)$. Moreover, for
all $\sigma\in B^{-1}(W)$ and index $i\in S_{2}\cup S_{3}$,
$\sigma[i]=ACK_{i-1}$. Also, for such a permutation $\sigma$, by
definition of graph $G$ it is easy to see that all pairs
$(i,\sigma[i])$ with $i\in S_{1}$ are edges in $G$. Hence $\sigma$
corresponds to a perfect matching in $G$.

Conversely, every perfect matching $M$ in $G$ naturally corresponds
to a sequence of pairs, that can be completed (by adding all pairs
$(i,ACK_{i-1})$ for all values $i$ not in $V_{1}$) to a mapping $A$
defined on $\{1,\ldots, n\}$. $A$ is actually a permutation. Indeed,
the values of parameter $ACK_{i}$, $i\in S_{2}\cup S_{3}$,  are all
different, and are not included in $V_{2}$. It follows that $A$ maps
$n$ numbers onto $n$ different numbers, hence it is a bijection.

To show that $A\in B^{-1}(W)$, assume that this was not the case,
and let $i$ be the smallest index such that $B_{A,i}\neq W_{i}$.
Thus $B_{A,i-1}=W_{i-1}$ where, by convention $B_{A,0}=0$.

{\bf Case 1} $B_{A,i}=B_{A,i-1}+1$. Since $W_{i}\neq B_{A,i}$ and
$W_{i}-W_{i-1}\leq 1$, the only possible alternatives are $W_{i} =
W_{i-1}$ or $W_{i} < W_{i-1}$. But then index $i$ is not in $V_{1}$
and is matched in $A$ to integer $ACK_{i-1}$. This contradicts the
assumption that $B_{A,i}=B_{A,i-1}+1$, since the packet with ID
$ACK_{i-1}$ is the first not received in the first $i-1$ phases, and
can thus be uploaded at stage $i$. The contradiction comes from our
assumption that sequences $B(A)$ and $W$ are different.

Similar arguments can be applied in the two remaining cases for the
evolution of sequence $B(A)$, and the conclusion of the argument is
that $A\in B^{-1}(W)$.
\end{proof}\qed

\begin{lemma}\label{graph}
Let $a_{1}\geq a_{2}\geq \ldots \geq a_{m}$ be the number of ones on
the first, second, $\ldots, m$'th row of $M_{G}$, the adjacency matrix
of $G$ (call $(a_{1}, \ldots, a_{m})$ {\em the type } of $M_{G}$).
Then we have
\begin{enumerate}

\item $G$ has a perfect matching if and only if for all $i=1,\ldots,
m$, $a_{i}\geq m+1-i$. When this condition holds, a perfect matching
in $G$ can be constructed by taking elements on the diagonal of
$M_{G}$.

\item The number of matchings in $G$ is given by
\begin{equation}\label{form}
|MATCH(G)|= a_{m}(a_{m-1}\cminus 1)(a_{m-2}\cminus 2)\cdot \ldots
 \cdot (a_{1}\cminus (m-1)).
\end{equation}

\end{enumerate}
\end{lemma}

\begin{proof}

Denote the cardinality of set $MATCH(G)$ by $\Gamma(a_{1}, \ldots,
a_{m})$ (to highlight its dependency on parameters $a_{1}, \ldots,
a_{m}$). Expand the permanent across the last row. Since $a_{1},
\ldots, a_{m-1}$ are all greater or equal to $a_{m}$, it follows that
$\Gamma(a_{1}, \ldots, a_{m})$ is the sum of the permanent of $a_{m}$
minors, all of them of type $(a_{1}\cminus 1, \ldots, a_{m-1}\cminus
1)$.  Thus, $ \Gamma(a_{1}, \ldots, a_{m})= a_{m}\cdot
\Gamma(a_{1}\cminus 1, \ldots, a_{m-1}\cminus 1)$, and
formula~(\ref{form}) immediately follows by noting that, for all
$i\geq 1$, $(a\cminus (i-1))\cminus 1 = a \cminus i$.

\end{proof} \qed

We now complete the proof of Theorem~\ref{main}.

\begin{enumerate}
\item Algorithm TwoStageGreedy in Figure~\ref{alg} produces a perfect
matching (if it exists). Its correctness follows from the recurrence
relations for parameter $ACK_{i}$ and Lemma~\ref{graph} (2). With a
little care the algorithm can be implemented in $O(|W|)$ time (using
$O(|W|)$ additional memory) as follows:
\begin{enumerate}

\item We use two buffers, $P$ and $Q$, each for $\lceil
\log_{2}(n)\rceil$ integers. They are intended to hold numbers $W_{i}$
and $W_{i-1}$. The for-loop can be implemented by simply scanning the
input from left to right, copying the correct information into buffers
$P$ and $Q$. Only two buffers are needed, provided we keep switching
roles of $P$ and $Q$ (they will alternately keep the last value
$W_{i}$). All test conditions in the algorithm involving these
numbers, as well as computing $W_{i}-W_{i-1}$, will be performed using
buffers $P$ and $Q$, and can be accomplished by scanning these buffers
$C$ times, for some fixed constant $C$.

\item The final for loop can be implemented in linear time by scanning
buffer $\sigma$ from left to right, using an additional counter for
the value of index $j$.

\item In the algorithm we keep incrementing several counters. The
problem of incrementing counters is well-known to have linear time
algorithms via amortized analysis \cite{CorLeiRivStein}.

\end{enumerate}

\item Computing $|MATCH(G)|$ using formula~(\ref{form}) can be done in
polynomial time as follows:
\begin{enumerate}

\item First, there is a {\em linear} time algorithm that, given input
$W$, outputs the list of numbers $a_{1}, \ldots, a_{m}$.

\item Given these numbers, computing $|MATCH(G)|$ can be accomplished
in time polynomial in $m+\lceil \log_{2}|MATCH(G)|\rceil$ by the
brute-force product computation in (\ref{form}). Since $|MATCH(G)|\leq
n!$ (simply because matchings correspond to permutations), it follows
by Stirling's approximation that $\lceil
\log_{2}|MATCH(G)|\rceil=O(n\log n)$. Thus, the running time is
polynomial in $|W|$.

\end{enumerate}

\begin{figure}
\dahntab{
\ \ \ \ \=\ \ \ \ \=\ \ \ \ \=\ \ \ \ \= \ \ \ \ \=\ \ \ \ \= \\
{\bf Algorithm TwoStageGreedy(W)} \\
\\
{\bf INPUT}: a vector $W=W_{1}\#W_{2}\#\ldots\#W_{n}\#\#$ of nonnegative integers. \\
\\
Let $\sigma$ be a vector of $n$ numbers of length $\lceil \log_{2}(n)\rceil$, initially all zero.\\
Let $ACK$ be a vector of $n+1$ numbers of length $\lceil \log_{2}(n)\rceil$, initially all zero,\\
with the exception of $ACK_0 = 1$.\\
Let $chosen$ be an $n$-bit vector, with all positions initially zero.\\
Let $W_0 = 0$.\\\\
for  $i = 1$ to  $n$ \\
\> \> if $(W_{i}-W_{i-1}>1)\OR ((i<n)\AND (W_{i}= 0)) \OR ((i=n)\AND (W_{i}\neq 0))$ \\
\> \> \> \keyw{reject} \\
\> \> else \\
\> \> \> if $(W_{i}=W_{i-1})$ \\
\> \> \> \> let $\sigma[i]=ACK_{i-1}$;  \\
\> \> \> \> let $chosen[ACK_{i-1}]=1$;  \\
\> \>  \> \> let $ACK_{i}=ACK_{i-1}+1$; \\
\> \> \> else \\
\> \> \> \> if $(W_{i}<W_{i-1})$ \\
\> \> \> \> \> let $\sigma[i]=ACK_{i-1}$;  \\
\> \> \> \> \> let $chosen[ACK_{i-1}]=1$;  \\
\> \>  \> \> \> let $ACK_{i}=ACK_{i-1}+W_{i}-W_{i-1}+1$; \\
\> \> \> \> else \\
\> \> \> \> \> /* $W_{i}=W_{i-1}+1$ */ \\
\> \> \> \> \> let $ACK_{i}=ACK_{i-1}$;  \\
\\
for  $i = 1$ to  $n$ \\
\> \> if $(\sigma[i]=0)$ \\
\> \> \> let $\sigma[i]=$ the first $j> ACK_{i-1}+1$ with $chosen[j]=0$;  \\
\\
\keyw{return} $\sigma$. \\
}
\caption{Algorithm for inverting buffer sequences}\label{alg}
\end{figure}

\end{enumerate}
\end{proof}\qed

The proof of Theorem~\ref{main} also implies that buffer equivalence
is a refinement of behavioral equivalence for permutations:

\begin{proposition}\label{refine}
Let $P$ and $Q$ be two {\em permutations} such that $P\equiv_{buf} Q$.
Then $P\equiv_{beh}Q$. \end{proposition}
\begin{proof}

Equations (\ref{eq4})--(\ref{eq6}) show that the value of parameter
$ACK_{i}$ can be recovered directly from the buffer sizes.  Since
$P$ and $Q$ are buffer equivalent, they have identical buffer size
sequences and, consequently, identical sequences of parameter
$ACK_{i}$. But it is easy to see that the sequence of packet IDs
(more precisely the corresponding sequence of byte IDs) ACKed by the
TCP protocol in the case of simple consecutive ACKs is precisely
$ACK_{i}$. Therefore $P$ and $Q$ are behaviorally equivalent.
\end{proof}\qed

\section{Reconstructing Packet Sequences with Repeats}

Buffer equivalence is not a refinement of behavioral equivalence in
the presence of repeats. The reason is that one cannot distinguish
between the case when the newly received packet is a repeat and Case
3 in the proof of Theorem~\ref{main} (in both cases the buffer size
stays the same). However, for a repeat packet the value of the $ACK$
parameter does not change, while for a packet in Case 3 the value of
the $ACK$ parameter increases by one.

One can modify the notion of buffer equivalence (in a somewhat
artificial way) to incorporate information whether the received
packet is a repeat or not. For instance, one can define $B_{A,i}$ to
be {\em minus} the buffer size when the $i$'th received packet is a
repeat. Denote this new mapping by $\overline{B}$.

\begin{definition}
Two sequences of packets $P$ and $Q$ are {\em modified buffer
equivalent} (written $P\equiv_{\overline{buf}} Q$) if
$\overline{B}(P)=\overline{B}(Q)$.
\end{definition}

The analog of Theorem~\ref{main} for mapping
$\overline{B}$ is

\begin{theorem}\label{main-2}
Let $W = W_1\#W_2\#\ldots\#W_n\#\#$ be a sequence of  integers.

Deciding whether $W$ is a valid buffer pattern, and in this case
constructing an ID sequence $A$ such that $A=\overline{B}^{-1}(W)$,
can be done in linear time. Counting the cardinality of the preimage
$\overline{B}^{-1}(W)$ can be done in polynomial time.

\end{theorem}

We only outline the proof, since it is very similar to that of
Theorem~\ref{main}. Given our use of negative numbers in the encoding,
we no longer have the positivity constraint for elements of
the candidate sequence $W$. However, we still require that only the
last element be zero.

The construction of graph $G$ is identical to that in the previous
case, since in all stages in $V_{1}$ we can guarantee that a new
packet is received. However, we  do {\em not} have a parsimonious
reduction of ID sequences to perfect matchings, since repeat packets
can complete a matching in $G$ in more than one way.

A polynomial-time counting algorithm exists, nevertheless, since we
can complement Lemma~\ref{graph} with

\begin{lemma}
We have
\begin{equation}
|\overline{B}^{-1}(W)|= |MATCH(G)| \times \left(\prod_{i\in R}
|W_{i}|\right),
\end{equation}
where $MATCH(G)$ is the set of all perfect matchings in $G$,
and $R=\{i\mbox{ }|W_{i}<0\}$, i.e. the set of stages in which a repeat packet
arrives. In
particular $\overline{B}^{-1}(W)\neq \emptyset$ if and only if $G$ has a
perfect matching.
\end{lemma}

Also, the construction shows that modified buffer equivalence {\em
is} a refinement of behavioral equivalence. Indeed, from the
sequence of modified buffer sizes one can uniquely reconstruct the
sequence of acknowledgments. The proof then proceeds just as the
proof of Proposition~\ref{refine}.

\section{Acknowledgments}

We acknowledge the anonymous referees of this paper for very useful
references and suggestions.

This work 
has been supported by the U.S.\ Department of
Energy under contracts W-705-ENG-36 and DE-AC52-06NA25396.
\vspace{-.2in}
\bibliographystyle{elsart-num}
\bibliography{/home/users/gistrate/bib/bibtheory}  
\end{document}